# New Quantum System of Wilson Orthogonal Polynomial

T. J. Taiwo

*Department of Mathematics, University of Lagos. Akoka. Lagos- State . P. O. Box 101017, Nigeria*

**Abstract**: We find a new quantum system associated with the Wilson Orthogonal Polynomial. In order to establish correspondence between the recent reformulation of quantum mechanics without potential function [1-2] and the convention quantum mechanics, we derived the potential function of the new quantum system using any of the proposed formula in [4]. To achieve this, we used the matrix elements of the potential function and the basis element of the configuration space.

## 1. Introduction

In paper [1] and [2], a reformulation of quantum mechanics without potential function was introduced. Basically, the essence of this reformulation was to enlarge the class solvable quantum systems. It was observed that there are quantum systems that could be described analytically but their potential function are either difficult to specify or impossible to realize analytically. For instance, the potential functions might be non-analytic, nonlocal, energy dependent, or the corresponding differential wave equation is higher second order and so on.

In brevity, the new reformulation implies that the wavefunction is to be expanded over a complete set of square – integrable basis functions in configuration space. That is (in atomic units $\hbar = m = 1$)

$$\psi(E, x) = \sqrt{\rho^\mu(\varepsilon)} \sum_n P_n^\mu(\varepsilon) \phi_n(x) \qquad (1)$$

This is assumed to be a bounded sum. $\{\phi_n(x)\}$ is the complete set of square- integrable basis functions in configuration space with coordinate $x$ and $P_n^\mu(\varepsilon)$ is the expansion coefficients which are orthogonal polynomials of order $n$ in the variable $\varepsilon$ which is some proper function of the energy. Also, $\mu$ represents a set of real parameters associated with the physical system and $\rho^\mu(\varepsilon)$ is the normalized weight function associated with the energy polynomials. The energy polynomials satisfy a symmetric three – term recursion relation

$$\varepsilon P_n^\mu(\varepsilon) = a_n^\mu P_n^\mu(\varepsilon) + b_{n-1}^\mu P_{n-1}^\mu(\varepsilon) + b_n^\mu P_{n+1}^\mu(\varepsilon), \qquad n = 1, 2, ..., \qquad (2)$$

where $P_0^\mu(\varepsilon) = 1$, $P_1^\mu(\varepsilon) = \alpha \varepsilon + \beta$ and $\{a_n^\mu, b_n^\mu\}$ are the recursion coefficients with $b_n^\mu \neq 0$ for all $n$. When $\alpha = 1/b_0^\mu$ and $\beta = -a_0^\mu / b_0^\mu$; $\{P_n^\mu(\varepsilon)\}$ is refer to as polynomial of the first kind. Also, the basis set $\{\phi_n(x)\}$ satisfy the boundary conditions and contains the kinematical information such as the angular momentum and the length scale. However, the detailed physical information about the system is contained in the energy polynomials $\{P_n^\mu(\varepsilon)\}$ and the corresponding weight function $\rho^\mu(\varepsilon)$. It is required the relevant energy polynomials to have

the following asymptotic (limits as $n \to \infty$)

$$P_n^\mu(\varepsilon) \approx n^{-\tau} A^\mu(\varepsilon) \times \cos\left[n^\xi \theta(\varepsilon) + \delta^\mu(\varepsilon)\right] \tag{3}$$

where $\tau$ and $\xi$ are real positive constants that depend on the particular polynomial. $A^\mu(\varepsilon)$ is the scattering amplitude and $\delta^\mu(\varepsilon)$ is the phase shift. Bound states, if they exist, occur at (infinite or finite) energies that make the scattering amplitude vanish. That is, the $m^{\text{th}}$ bound state occurs at an energy $E_m$ such that $A^\mu(\varepsilon) = 0$ and the corresponding bound state is written as

$$\psi(E_m, x) = \sqrt{\omega^\mu(\varepsilon_m)} \sum_n Q_n^\mu(\varepsilon_m) \phi_n(x) \tag{4}$$

where $\{Q_n^\mu(\varepsilon)\}$ are the discrete version of the polynomials $\{P_n^\mu(\varepsilon)\}$ and $\omega^\mu(\varepsilon_m)$ is the associated discrete weight function. Hence, without potential function, the physical properties of the system are deduced from the properties of the associated orthogonal polynomials. Such properties include the shape of the weight function, nature of the generating function, distribution and density of the polynomial zeros, recursion relation, asymptotic, differential or difference equations, etc.

As a result of this reformulation, a new quantum system – **Wilson- Racah Quantum System** was obtained in [3]. The discrete energy spectrum, scattering phase shift, and discrete wavefunction of this quantum system were obtained and analyzed.

In order to establish a correspondence with the conventional formulation of quantum mechanics, procedures for reconstructing the potential function numerically in this new reformulation were formulated in [4]. It was shown that once the potential matrix element of the potential function is gotten and with the basis element; the potential function can be derived using any of the four formulas. Although, each of the four formulas gives different degree of accuracy.

So in this paper, our aim is to get another quantum system of the Wilson Orthogonal polynomial. In section 2, we show how the energy spectrum, scattering phase shift, and wavefunction of the new quantum system can be derived. In section 3, we derived the potential matrix elements of the new quantum system, and a plot of the potential function will be given.

## 2. The New Quantum System

In [3-Appendix B], from the asymptotic $(n \to \infty)$ of the Wilson polynomial, the scattering phase shift for any associated physical system was derived as

$$\delta(\varepsilon) = \arg\left[\Gamma(2iy)/\Gamma(\mu + iy)\Gamma(v + iy)\Gamma(a + iy)\Gamma(b + iy)\right] \tag{5}$$

where $y = \varepsilon(E)$ such that $y \geq 0$. For this new quantum system, we choose $\varepsilon = \lambda/k$, where $E = \frac{1}{2}k^2$ and $\lambda^{-1}$ is the length scale of the system in atomic units $\hbar = m = 1$, then $y = \varepsilon = \lambda/\sqrt{2E}$. If all the parameters of the Wilson polynomial are positive, there will be no bound states (that is continuous scattering states) while if $\mu < 0$ and $\mu + v$, $\mu + a$, $\mu + b$ are positive then there co-exit the continuum scattering states and $N$ bound states, where $N$ is the largest integer less than or equal to $-\mu$. Now from the zero of the scattering amplitude $A(\varepsilon)$ [3-Appendix B] dictates that the bound states occur at energies $\{\varepsilon_m\}_{m=0}^N$ such that $iy = -(m + \mu)$ giving the bound states energy spectrum as

$$E_m = -\frac{\lambda^2}{2(m + \mu)^2} \tag{6}$$

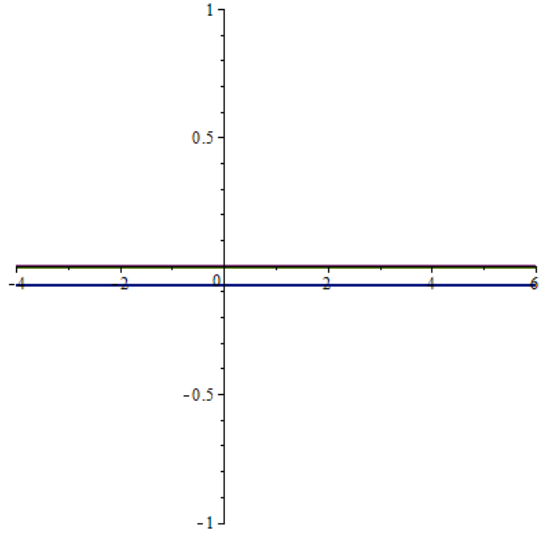

**FIG.1**. The bound states energy spectrum (6) for the new quantum system with physical parameters: $\lambda = 0.2$, $\mu = -0.5$, and $m = 0,1,2,...,$

The total wavefunction for the continuous energy $\varepsilon$ and discrete energy $\varepsilon_m$ can be written as

$$\psi_m(E,x) = \sqrt{\rho^\mu(\varepsilon)} \sum_{n=0}^{\infty} W_n^\mu(\varepsilon^2;v;a,b)\phi_n(x) + \sqrt{\rho_m^N(\varepsilon)} \sum_{n=0}^{N} W_n^\mu(-(m+\mu)^2;v;a,b)\phi_n(x) \qquad (7)$$

where $\rho^\mu(\varepsilon)$ and $\rho_m^N(\varepsilon)$ are the continuous and discrete normalized weight function from the augmented continuous orthogonality relation of the Wilson polynomial (See Equation 7 in [3]). $W_n^\mu(\varepsilon^2;v;a,b)$ is the Wilson orthogonal polynomial. Our basis function is chosen in one dimensional configuration space with coordinate $-\infty < x < \infty$ as $\phi_n(x) = [\pi 2^{n+m} n!]^{-1/2} e^{-\lambda^2 x^2/2} H_n(\lambda x)$, where $H_n(z)$ is the Hermite Polynomial of degree $n$ in $z$. .

## 3. Potential Function of New Quantum System

From conventional quantum mechanics, the Hamiltonian, $H = T + V$, is sum of the kinetic energy operator $T$ and the potential function $V$. Therefore, we have $V = H - T$. Now to get the potential matrix elements in the basis $\{\phi_n(x)\}$, we need first to get the matrix elements of the Hamiltonian operator $H$ and kinetic energy operator both in this basis. Since $T = -\frac{1}{2}\frac{d^2}{dx^2}$, in one dimension coordinate $x$, and also given as $T = -\frac{1}{2}\frac{d^2}{dr^2} + \frac{\ell(\ell+1)}{2r^2}$, in three dimensions with spherical symmetry and radial coordinate $r$ ($\ell$ is the angular momentum quantum number), its matrix representation could be easily derived by operating $T$, the kinetic operator, on the basis elements $\{\phi_n(x)\}$.

Since the total wave function of the system is $\Psi(t,x) = e^{-iEt/\hbar}\psi(E,x)$; so we have $H\Psi = i\hbar\frac{\partial}{\partial t}\Psi = E\Psi$. Hence, we can therefore write the wave operation as

$$H|\psi\rangle = E|\psi\rangle \qquad (8)$$

$H$ is the Hamiltonian operator and $E$ is the energy of the quantum system.

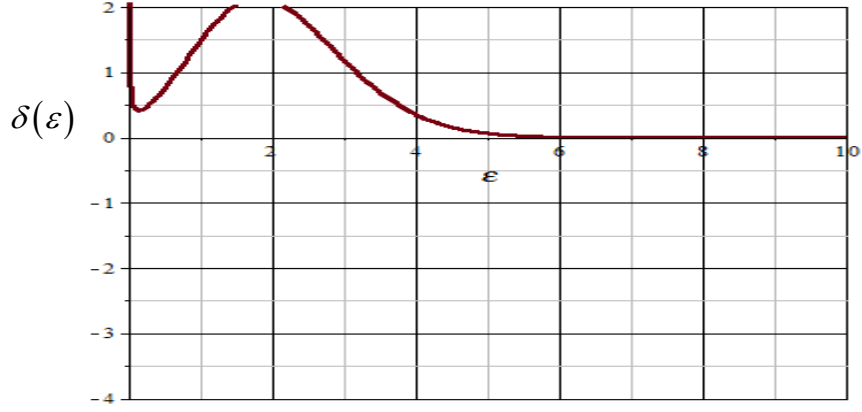

**FIG.2.** The scattering phase shift (in units of $\pi$) for the new quantum system with physical parameters: $\lambda = 0.2$, $a = 0.6$, $b = 0.5$, $\mu = 0.8$, $\nu = 0.3$. The energy variable is $\varepsilon = \lambda/k$, where $E = \frac{1}{2}k^2$ and $\lambda^{-1}$ is the system's length scale.

Using (1) in (8) and projecting from the left by $\langle \phi_n |$ gives the equation $\sum_m P_m^\mu \langle \phi_n | H | \phi_m \rangle = E \sum_m P_m^\mu \langle \phi_n | \phi_m \rangle$. We can rewrite this as a generalized eigenvalue matrix equation

$$\tilde{H} | P \rangle = E \Omega | P \rangle \qquad (9)$$

where $\tilde{H}$ is the matrix representation of the Hamiltonian operator in the basis $\{\phi_n(x)\}$ and $\Omega$ is the overlap matrix of the basis elements, $\Omega_{n,m} = \langle \phi_n | \phi_m \rangle$ (i.e., matrix representation of the identity). Now, since the energy polynomial satisfies (2); this can be rewritten as in matrix form as $\Sigma | P \rangle = \varepsilon | P \rangle$, where $\Sigma_{n,m} = a_n^\mu \delta_{n,m} + b_{n-1}^\mu \delta_{n,m+1} + b_n^\mu \delta_{n,m-1}$ is the tridiagonal symmetric matrix of the three term recursion relation in (2). Therefore, the wave equation (6) is equivalent to three term recursion relation of the energy polynomials $\{P_n^\mu(\varepsilon)\}$ and implies that the wave operator matrix $J = \tilde{H} - E\Omega$ will be tridiagonal and symmetric. By this, the matrix representation of Hamiltonian operator can be easily gotten from the three term recursion relation of the energy polynomial and the potential matrix elements of the potential function are easily obtained too.

Since, we always require the matrix wave operator $J = V + T - E\Omega$ be tridiagonal; but if $\Omega$ is non-tridiagonal (i.e., the basis elements are neither orthogonal nor tri-thogonal), then the kinetic energy matrix $T$ will have corresponding energy –dependent components to cancel out the non tridiagonal components. After this, if there still exist further non-tridiagonal components then they will be eliminated by the counter components in $V$. Now we find the potential function that defined the wavefunction (7)

Operating the kinetic operator $T$ on the basis element gives

$$T_{n,m} = -\frac{1}{2}\langle \phi_n | \frac{d^2}{dx^2} | \phi_m \rangle = \lambda^2 \left( n + \frac{1}{2} \right) \delta_{n,m} - \frac{1}{2}\lambda^2 \langle n | (\lambda x)^2 | m \rangle \qquad (10)$$

where $\langle n | f(z) | m \rangle = \left[ \pi 2^{n+m} n! m! \right]^{-1/2} \int_{-\infty}^{+\infty} e^{-z^2} f(z) H_n(z) H_m(z) dz$. Since (10) will not be tridiagonal due to the last expression in R.H.S, then we eliminate it by a counter - term in the sought after potential function. So, it is eliminated by the term $+\frac{1}{2}\lambda^4 x^2$, which is a harmonic oscillator potential term. Hence, the potential function we are looking for is

$$V(x) = \frac{1}{2}\lambda^4 x^2 + \tilde{V}(x) \tag{11}$$

The unknown potential function $\tilde{V}(x)$ associated with first term in (10) will be resolved numerically. Using the recursion relation of the Hermite polynomial in (10) without the last term

$$T_{n,m} = -\frac{1}{2}\langle\phi_n|\frac{d^2}{dx^2}|\phi_m\rangle = \frac{\lambda^2}{4}\left[(2n+1)\delta_{n,m} - \sqrt{n(n-1)}\delta_{n,m+2} - \sqrt{(n+1)(n+2)}\delta_{n,m-2}\right] \tag{12}$$

which is tridiagonal in function space with only odd and or only even indices. The Hamiltonian matrix element is gotten by using the following parameters: $y = \lambda/\sqrt{2E}$, $\mu = \nu$, and $a = b$ in the three term recursion relation of the Wilson orthogonal polynomial (A7)

$$H_{n,m} = \frac{1}{\lambda^2}\left[\left(n+\mu+a-\frac{1}{2}\right)^2 - \left(\mu-\frac{1}{2}\right)^2 - \left(a-\frac{1}{2}\right)^2 + \frac{1}{4}\right]\delta_{n,m} \tag{13}$$

$$-\frac{1}{2\lambda^2}\left\{(n+\mu+a-1)\sqrt{\frac{n(n+2\mu-1)(n+2a-1)(n+2\mu+2a-2)}{(n+\mu+a-1)^2-\frac{1}{4}}}\delta_{n,m+1} + (n+\mu+a)\sqrt{\frac{(n+1)(n+2\mu)(n+2a)(n+2\mu+2a-1)}{(n+\mu+a)^2-\frac{1}{4}}}\delta_{n,m-1}\right\}$$

Hence the matrix elements of the potential function $\tilde{V}(x)$ is $\tilde{V} = H - \tilde{T}$. Finally fig.3 is the full plot of the potential function (11) using the first two numerical formulas given in [4] for getting the potential functions:

$$(a) \quad V(x) \cong \frac{\sum_{n=m=0}^{N-1} V_{nm}\bar{\phi}_n(x)\bar{\phi}_m(x)}{\sum_{n=0}^{N-1}\bar{\phi}_n(x)\bar{\phi}_n(x)} \qquad (b) \quad V(x) \cong \sum_{m=0}^{N-1}\frac{\bar{\phi}_m(x)}{\bar{\phi}_0(x)}V_{m,0} \tag{14}$$

Detailed analysis and derivation of (14) had been given in [2] and [4]. We obtained a more accurate and stable result using (14a) since all the full matrix elements $\frac{1}{2}N(N+1)$ of the potential function were used compared to (14b) which has $N$ elements. Note the order of the matrix is $N \times N$.

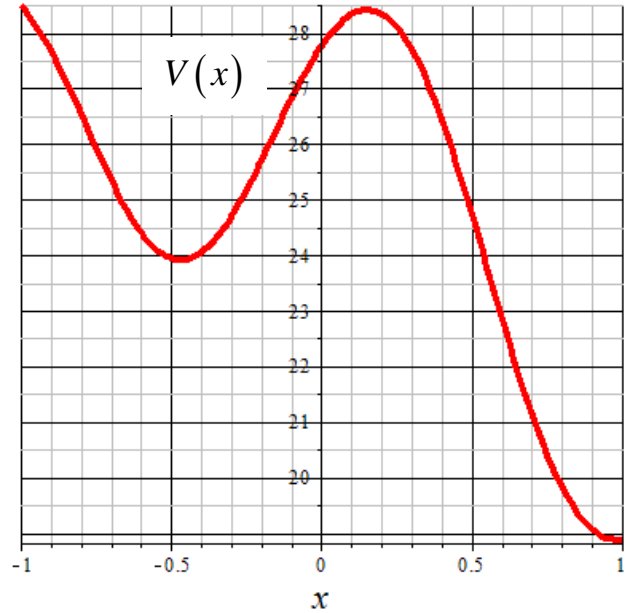

**FIG.3**. The potential function (11) computed using (14) with physical parameters: $\lambda = 0.5$, $a = 0.6$, $b = 0.5$ $\mu = 0.8$, $\nu = 0.3$.

## 4. Conclusion

In this paper, we derived analytically a new quantum system associated with the Wilson Orthogonal polynomial. The energy spectrum, scattering phase shift, and wavefunction of the system were shown. In an effort to establish the correspondence between recent reformulation of quantum mechanics without potential function and the convention method; we were able to graphically get the potential function of the new quantum system using two of the formulas given in [4].

## Acknowledgement

The author highly appreciates the support of the Saudi Centre for Theoretical Physics during the progress of this work. Specifically, this paper is dedicated to **Prof. A.D. Alhaidari** for his fatherly role and support in impacting knowledge. Allah in His infinite mercy will continue to bless him more.

## Appendix A: The Wilson Orthogonal Polynomial

The Wilson polynomial, $\tilde{W}_n^\mu(y^2;v;a,b)$ can be defined

$$\tilde{W}_n^\mu(y^2;v;a,b) = \frac{(\mu+a)_n(\mu+b)_n}{(a+b)_n n!} {}_4F_3\left(\begin{array}{c}-n, n+\mu+v+a+b-1, \mu+iy, \mu-iy \\ \mu+v, \mu+a, \mu+b\end{array}\bigg|1\right) \quad (A1)$$

where ${}_4F_3\left(\begin{array}{c}a,b,c,d \\ e,f,g\end{array}\bigg|z\right) = \sum_{n=0}^{\infty} \frac{(a)_n(b)_n(c)_n(d)_n}{(e)_n(f)_n(g)_n} \frac{z^n}{n!}$ is the Hypergeometric function and

$(a)_n = a(a+1)(a+2)...(a+n-1) = \frac{\Gamma(n+a)}{\Gamma(a)}$. The generating function of these polynomials is

$$\sum_{n=0}^{\infty} \tilde{W}_n^\mu(y^2;v;a,b) t^n = {}_2F_1\left(\begin{array}{c}\mu+iy, v+iy \\ \mu+v\end{array}\bigg|t\right) {}_2F_1\left(\begin{array}{c}a-iy, b-iy \\ a+b\end{array}\bigg|t\right) \quad (A2)$$

Their three term recursion relation $(n=1,2,3...,)$ is

$$y^2 \tilde{W}_n^\mu = \left[\frac{(n+\mu+v)(n+\mu+a)(n+\mu+b)(n+\mu+v+a+b-1)}{(2n+\mu+v+a+b)(2n+\mu+v+a+b-1)} + \frac{n(n+v+a-1)(n+v+b-1)(n+a+b-1)}{(2n+\mu+v+a+b-1)(2n+\mu+v+a+b-2)} - \mu^2\right]\tilde{W}_n^\mu$$
$$-\frac{(n+\mu+a-1)(n+\mu+b-1)(n+v+a-1)(n+v+b-1)}{(2n+\mu+v+a+b-1)(2n+\mu+v+a+b-2)}\tilde{W}_{n-1}^\mu - \frac{(n+1)(n+\mu+v)(n+a+b)(n+\mu+v+a+b-1)}{(2n+\mu+v+a+b)(2n+\mu+v+a+b-1)}\tilde{W}_{n+1}^\mu \quad (A3)$$

The initial seeds $(n=0)$ for this recursion are $\tilde{W}_0^\mu = 1$ and $\tilde{W}_1^\mu = \frac{(\mu+a)(\mu+b)}{(a+b)} - \frac{\mu+v+a+b}{(\mu+v)(a+b)}(y^2+\mu^2)$.

The orthogonality relation of the polynomial is

$$\frac{1}{2\pi}\int_0^\infty \frac{\Gamma(\mu+v+a+b)|\Gamma(\mu+iy)\Gamma(v+iy)\Gamma(a+iy)\Gamma(b+iy)|^2}{\Gamma(\mu+v)\Gamma(a+b)\Gamma(\mu+a)\Gamma(\mu+b)\Gamma(v+a)\Gamma(v+b)|\Gamma(2iy)|^2}\tilde{W}_n^\mu(y^2;v;a,b)\tilde{W}_m^\mu(y^2;v;a,b)dy$$
$$= \left(\frac{n+\mu+v+a+b-1}{2n+\mu+v+a+b-1}\right)\frac{(\mu+a)_n(\mu+b)_n(v+a)_n(v+b)_n}{(\mu+v)_n(a+b)_n(\mu+v+a+b)_n n!}\delta_{nm} \quad (A4)$$

The normalized weight function is

$$\rho^\mu(y;v;a,b) = \frac{1}{2\pi}\frac{\Gamma(\mu+v+a+b)|\Gamma(\mu+iy)\Gamma(v+iy)\Gamma(a+iy)\Gamma(b+iy)/\Gamma(2iy)|^2}{\Gamma(\mu+v)\Gamma(a+b)\Gamma(\mu+a)\Gamma(\mu+b)\Gamma(v+a)\Gamma(v+b)} \quad (A5)$$

Finally, the orthonormal version of this polynomial is

$$W_n^\mu(y^2;v;a,b) = \sqrt{\left(\frac{2n+\mu+v+a+b-1}{n+\mu+v+a+b-1}\right)\frac{(\mu+v)_n(a+b)_n(\mu+v+a+b)_n n!}{(\mu+a)_n(\mu+b)_n(v+a)_n(v+b)_n}} \tilde{W}_n^\mu(y^2;v;a,b)$$

$$= \sqrt{\left(\frac{2n+\mu+v+a+b-1}{n+\mu+v+a+b-1}\right)\frac{(\mu+v)_n(\mu+a)_n(\mu+b)_n(\mu+v+a+b)_n}{(a+b)_n(v+a)_n(v+b)_n n!}} \,{}_4F_3\!\left(\begin{array}{c}-n, n+\mu+v+a+b-1, \mu+iy, \mu-iy \\ \mu+v, \mu+a, \mu+b\end{array}\bigg|1\right) \quad \text{(A6)}$$

The three – term recursion relation for the orthonormal version is

$$y^2 W_n^\mu = \left[\frac{(n+\mu+v)(n+\mu+a)(n+\mu+b)(n+\mu+v+a+b-1)}{(2n+\mu+v+a+b)(2n+\mu+v+a+b-1)} + \frac{n(n+v+a-1)(n+v+b-1)(n+a+b-1)}{(2n+\mu+v+a+b-1)(2n+\mu+v+a+b-2)} - \mu^2\right] W_n^\mu$$

$$-\frac{1}{2n+\mu+v+a+b-2}\sqrt{\frac{n(n+\mu+v-1)(n+a+b-1)(n+\mu+a)(n+\mu+b-1)(n+v+a-1)(n+v+b-1)(n+\mu+v+a+b-2)}{(2n+\mu+v+a+b-3)(2n+\mu+v+a+b-1)}} W_{n-1}^\mu$$

$$-\frac{1}{2n+\mu+v+a+b}\sqrt{\frac{(n+1)(n+\mu+v)(n+a+b)(n+\mu+a)(n+\mu+b)(n+v+a)(n+v+b)(n+\mu+v+a+b-1)}{(2n+\mu+v+a+b-1)(2n+\mu+v+a+b-1)}} W_{n+1}^\mu \quad \text{(A7)}$$